\documentclass[useAMS,usenatbib]{mn2e}
\usepackage{psfig}
\usepackage{epsfig}
\bibpunct{(}{)}{;}{a}{}{,}

\begin{document}
\title {The coherence of kHz quasi-periodic oscillations in the X-rays from accreting neutron stars} \author[Barret, Olive, Miller]{Didier Barret$^{1}$,
  Jean-Francois Olive$^{1}$, \& M. Coleman Miller$^{2}$
  \thanks{E-mail: Didier.Barret@cesr.fr}\\
$^1$Centre d'Etude Spatiale des Rayonnements, CNRS/UPS, 9 Avenue du
Colonel Roche, 31028 Toulouse Cedex 04, France\\
$^2$Department of Astronomy, University of Maryland, College Park, MD
20742-2421, United States}
\date{Accepted 2006 May 16. Received 2006 May 16; in original form 2006 March 7 }

\pagerange{\pageref{firstpage}--\pageref{lastpage}} \pubyear{2006}
\maketitle

\label{firstpage}

\begin{abstract}
We study in a systematic way the quality factor of the lower and upper
kHz QPOs in a sample of low luminosity neutron star X-ray binaries, showing both QPOs varying over a
wide frequency range. The sample includes 
4U~1636--536, 4U~1608--522, 4U~1735--44,  4U~1728--34, 4U~1820--303 and 
4U~0614+09. We find that all sources except 4U~0614+091 show evidence of a drop in the quality
factor of their lower kHz QPOs at high frequency. For 4U~0614+091 only the rising part of the quality factor versus frequency curve has been sampled so far. At the same time, in all sources but 4U~1728--34, the quality factor of the upper kilo-Hz QPO increases all the way to the highest detectable frequencies. We show that the high-frequency behaviours of
both the lower and upper kHz QPO quality factors are consistent with
what is expected if the drop is produced by the approach of an active
oscillating region to the innermost stable circular orbit: the existence of which is a key feature of General Relativity in the strong field regime. Within this interpretation, our results imply gravitational masses around $2 M_\odot$ for the neutron stars in those systems.

\end{abstract}


\section{Introduction}
Kilohertz Quasi-Periodic Oscillations (kHz QPOs) have now been
detected by the Rossi X-ray Timing Explorer (RXTE, Bradt et al. 1993) from about 25
neutron star low-mass X-ray binaries (van der Klis 2006 and references therein). These signals, whose
timescales correspond to the dynamical times of the innermost
regions of the accretion flow, have triggered much interest because they 
may carry imprints of strong field general relativity,
such as the existence of an innermost stable circular orbit
(ISCO) around a sufficiently compact neutron star. For instance, a
drop in the amplitude and quality factor of the QPOs at some
limiting frequency was proposed as a possible signature of the
ISCO (e.g., Miller, Lamb, \& Psaltis 1998).

Following this idea, we studied in a systematic way the QPO
properties of 4U~1636--536, with emphasis on the dependency of
the quality factor ($Q\equiv \nu/{\rm FWHM}$) and amplitude with 
frequency (Barret, Olive, \& Miller 2005a,b). We have
thus shown that the lower and upper kHz QPOs of 4U~1636--536
follow two distinct tracks in a Q versus frequency plot. The
quality factor of the lower kHz QPO increases with frequency
up to a maximum of $Q\sim 200$ at $\sim 850$ Hz, then drops
precipitously to $Q\sim 50$ at the highest detected
frequencies $\nu_{\rm lower}\sim 920$~Hz (a similar behaviour was seen earlier from 4U1608-52, Barret et al. 2005c) On the other hand,
the quality factor of the upper kHz QPO increases steadily all
the way to the highest detectable QPO frequency, although the
quality factor is lower than for the lower QPO.  The rms
amplitudes of both the upper and lower kHz QPOs decrease
steadily towards higher frequencies.  

In this paper, we extend
the analysis carried out for 4U~1636--536 to all low-luminosity
neutron star binaries for which kHz QPOs have been detected over a
large range of frequencies. The sample includes: 
4U~1636--536, 4U~1608--522, 4U~1820--303, 4U~1735--44, 4U~1728--34,
4U~0614+09. An exhaustive list of references on QPOs from these sources is available in van der Klis (2006). In \S~2 we describe our analysis procedure, and in \S~3
we discuss the implications of our results and show how the high-frequency
dropoff in $Q$ can be accommodated quantitatively in a toy model
based on the approach to the ISCO.

\section{Data analysis}

For the purpose of this paper, we have retrieved all science event 
files from the RXTE archives up to the end of 2004 for all six 
sources. The files are identified with their observation identifier
(obs ID) following the RXTE convention. An Obs ID identifies a
temporally contiguous collection of data from a single pointing.
Only files with time resolution better than or equal to 250
micro-seconds and exposure times larger than 600 seconds are
considered. No filtering on the raw data is performed, which means
that all photons are used in the analysis.  Type I X-ray bursts and
data gaps are removed from the files.
\begin{figure*} 
\includegraphics[width=0.8\textwidth]{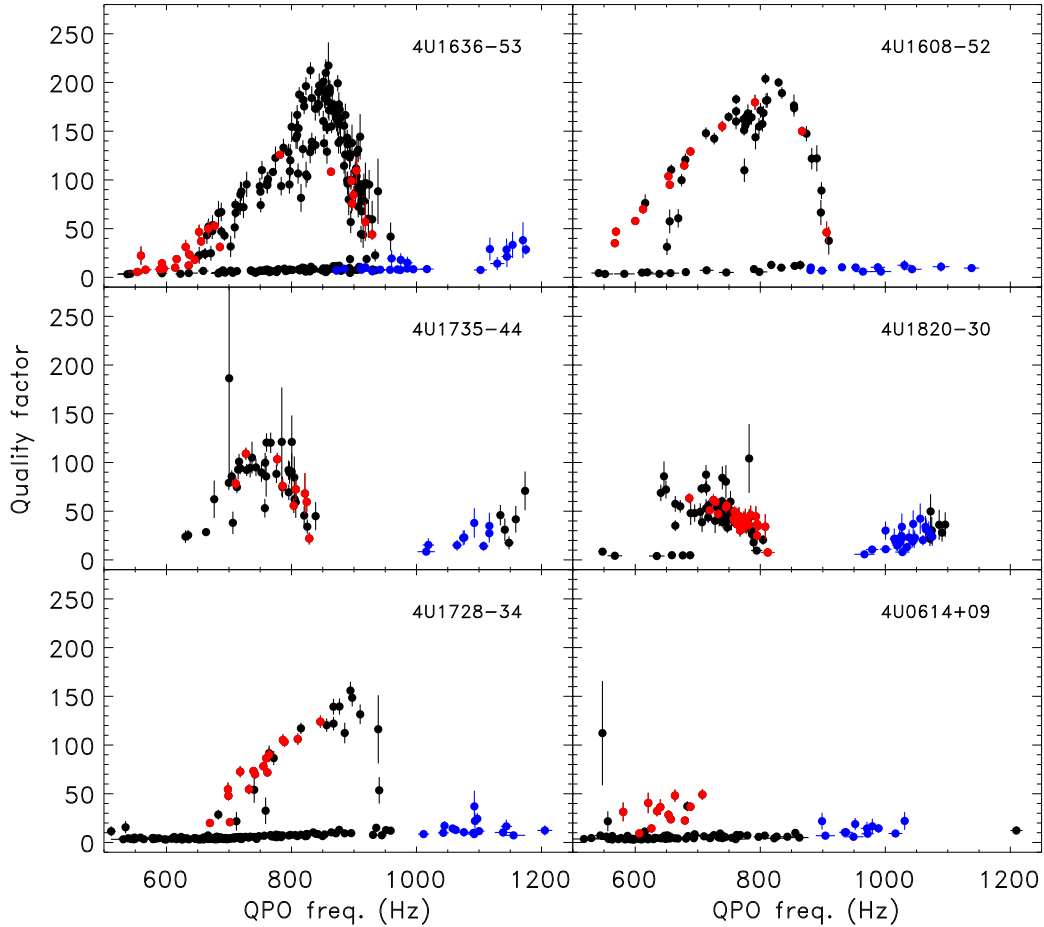}
\caption[]{Quality factor versus frequency of all fitted QPOs:
red circles identify fitted lower QPOs, blue circles identify
fitted upper QPOs, black circles correspond to single QPOs. Each data point represents the average over a continuous data segment. }
\label{barret_f0}
\end{figure*}

For each file identified with its Obs ID, we have computed Leahy normalized Fourier power  
density spectra (PDS) between 1 and 2048 Hz over 8~s intervals (with  
a 1 Hz resolution). $N$ 8-second PDS are thus computed. $N$ is typically around 400 in most files, whose duration $\sim 3200$ is consistent with the orbital period of the RXTE spacecraft. A Fourier power spectrum averaging the $N$ PDS 
is first computed. This averaged PDS is then searched for a QPO  
using a scanning technique which looks for peak excesses above the  
Poisson counting noise level (Boirin et al. 2000). No fit is performed
at this stage as the scanning procedure returns the centroid frequency of each excess and an approximation of its spread in frequency (a rough measure of its total width). In
case of the presence of two excesses, the one with the higher  
significance is considered: its centroid frequency is $\nu_0$ and its spread is $w_0$.

We now wish to attribute to all $N$ PDS the best estimate of the instantaneous QPO frequency $\nu_{qpo_{\rm 1\dots N}}$ to apply later on the shift-and-add technique. To start, $\nu_{qpo_{\rm 1\dots N}}$ is set to  $\nu_0$. We define a window of 25 Hz on both sides of the excess centered at $\nu_0$ and recursively search for excesses over time intervals of shorter and shorter durations, i.e. in a smaller and smaller number of 8 second PDS averages. More specifically, in the first iteration, we thus consider two consecutive time intervals averaging the 8-sec PDS from number 1 to $N/2$  to produce PDS$_{1,1}$ and the 8-sec PDS from number $N/2+1$ to $N$ to produce PDS$_{1,2}$. In PDS$_{1,1}$ and PDS$_{1,2}$, we search for peak excesses between $\nu_0 - w_0/2 - 25$~Hz and $\nu_0 +w_0/2+ 25$~Hz, still above the Poisson counting noise level at the 4$\sigma$ level. For instance, if an excess is detected PDS$_{1,1}$ at a frequency $\nu_{1,1}$ (spread $w_{1,1}$), then we set $\nu_{qpo_{\rm 1\dots N/2}}$ to $\nu_{1,1}$ before the next iteration. If on the other hand, no excess is detected in PDS$_{1,2}$, then $\nu_{qpo_{\rm N/2+1\dots N}}$ remains unchanged at $\nu_0$. In the next iteration, we repeat the procedure considering two further intervals averaging the 8-sec PDS from number 1 to $N/4$  to produce PDS$_{2,1}$ and the 8-sec PDS from number $N/4+1$ to $N/2$ to produce PDS$_{2,2}$, and search in both PDS an excess between $\nu_{1,1} - w_{1,1}/2-25$~Hz and $\nu_{1,1} +w_{1,1}/2+ 25$~Hz. The tracking procedure stops when no more excess is detected in any of the intervals analyzed. At the end, the procedure thus returns the best possible estimate of the instantaneous QPO frequency in each 8 second PDS. We have checked through simulations of a QPO signal of varying frequency and of amplitude similar to the real data that this procedure follows with great accuracy the changes in frequency. 

In order to reduce the contribution of the long term frequency drift to the broadening of the QPO profile, within a data file, we then shift-and-add (M\'endez et al. 1998) the individual 8 second PDS to the mean QPO frequency over the file. The shifted and added PDS  is then searched for QPOs with the scanning routine. A quality check is performed at this stage with the scanning routine to verify that the tracked QPO in the resulting PDS has a centroid frequency consistent with $\nu_0$ and a spread smaller than $w_0$. Either one or two QPOs are then fitted with a  Lorentzian of three parameters (frequency, full width at half maximum, and normalization) to which a constant is added to account  for the counting noise level (close to 2.0 in a Leahy normalized PDS). The fitting is carried with  the  XSPEC 11.3.2 spectral
package (Arnaud 1996), taking advantage of the robustness of its
fitting procedures  (including the error computations). Next, we
keep QPOs which are detected above $2.5\sigma$ (the significance
being defined as the integral of the Lorentzian divided by its
error), above 500 Hz, with a quality factor larger than 3. 

In Figure 1, we show the quality factor of the QPOs detected for all
six sources. Although with fewer details than for 4U~1636--536, the
same trends are present for all the other sources as well. In particular a sharp drop of the quality
factor of the lower kHz QPO is also seen in 4U~1608--522 and
4U~1728--34. The drop is also seen in 4U~1735--44 and to a lesser
extent in 4U~1820--303, although the maximum quality factor reached
is significantly lower than for the previous sources. For 4U~0614+09, whereas the rise of $Q$ is clearly suggested, the drop cannot be inferred. In all sources, but possibly 4U~1728--34, as shown at the bottom of the six panels, the quality factor of
the upper kHz QPOs shows a positive correlation with frequency all
the way to the highest detectable frequency.

As an indication, in those files in which the QPO tracked is detected in at least five intervals, the mean tracking timescales is $\sim 230, 70, 180, 220, 140$ and $290$ seconds for 4U~1636--536, 4U~1608--522, 4U~1735--44,  4U~1728--34, 4U~1820--303 and 4U~0614+09, respectively. These averages are only representative because they include files in which the lower QPO was followed and files in which the upper QPO was followed by the tracking procedure, and it is known that the tracking timescales is longer for the upper QPO than for the lower QPO, and that for the lower QPO the tracking timescales depend on $Q$; the higher $Q$ the smaller the tracking timescales (see Fig 1 in Barret et al. 2005a). This explains why the mean tracking timescale for 4U0614+091 is the longest because first, there are many files in which the upper QPO was followed by the tracking procedure and second only the low $Q$ domain of the lower QPO has been sampled so far with RXTE. One can also estimate the minimum tracking timescale for the lower QPO. It is 24, 24, 48, 56, 40, and 168 seconds for 4U~1636--536, 4U~1608--522, 4U~1735--44,  4U~1728--34, 4U~1820--303 and 4U~0614+09 respectively.

Using Figure 1, it is easy to determine which of the QPOs was followed
in the data file by the tracking procedure (even when a single QPO was detected). In order to get a better determination of the
dependency of the quality factor of the two QPOs with frequency, we
identify those data files in which the QPO tracked by the procedure
described above is the lower and upper QPO respectively. The frequency
range covered by the lower and upper QPO are then divided in intervals
of 25 Hz and 50 Hz respectively. All the 8 second PDS with a tracked
QPO falling into a frequency bin are then shifted to the central bin
frequency. The QPOs in the resulting PDS are then fitted. The results
of the fits are presented in Figure 2 and 3 for the quality factor and
amplitudes of both QPOs.  The trends seen in Figure 1 are then much
clearer (this is obvious in the case of 4U~1735--44). For 4U~1820--303, only the decaying part of the quality factor
versus frequency curve has been sampled. The behavior of the upper QPO is strikingly different from the lower QPO one. In all, but 4U~1728--34, the quality factor increases all the way to the highest detectable frequencies. In 4U~1728--34, the peak of the quality factor is reached at $\sim 1050$ Hz, beyond which a significant decrease is observed. In all sources, for the lower QPO, there is a clear trend for the amplitude to increase, then to decrease with frequency. The tendency for the amplitude of the upper QPO to decrease with frequency is also present, indicating that the quality factor and amplitude of the upper QPO anticorrelates towards its high-frequency end. Comparing Figure 2 and 3, a positive correlation between the quality factor and amplitude of the lower QPO can be inferred during the rise of its quality factor.

In Figure 4, we show the frequency difference between the two QPOs
against the lower QPO frequency. The frequency at which the
measured quality factor reached its maximum is also indicated,
together with the spin frequency or half the spin frequency
when available. This figure indicates that the peak of the quality factor of the lower QPO is not connected to the frequency difference between the two QPOs being equal to the spin frequency of the neutron star.

\begin{figure} 
\includegraphics[width=0.45\textwidth]{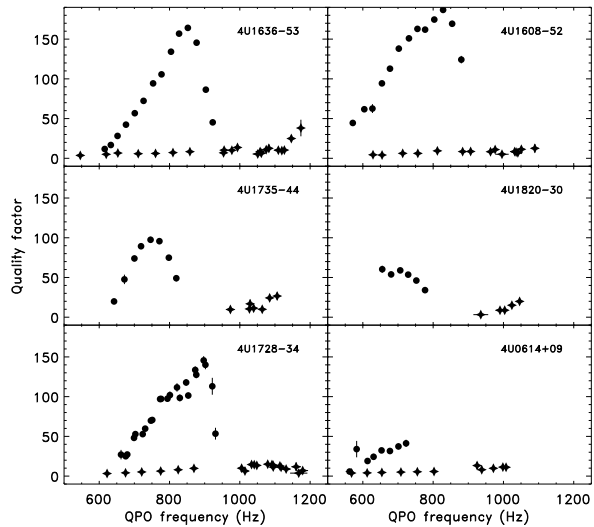}
\caption[]{Quality factor versus frequency of the lower (filled circles)
and upper QPOs (filled stars). The lower QPO has been sampled with a frequency bin of 25 Hz, whereas the bin is 50 Hz for the upper QPO.}
\label{barret_f1}
\end{figure}

\begin{figure} 
\includegraphics[width=0.45\textwidth]{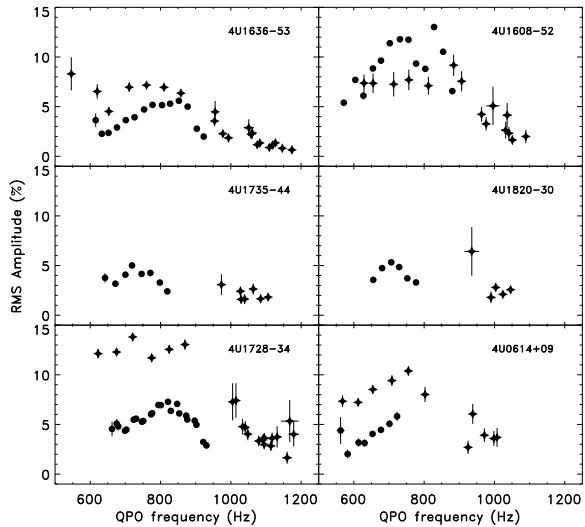}
\caption[]{Same as Figure 2. RMS amplitude versus frequency of the lower (filled circles)
and upper QPOs (filled stars).}
\label{barret_f2}
\end{figure}

\begin{figure} 
\includegraphics[width=0.45\textwidth]{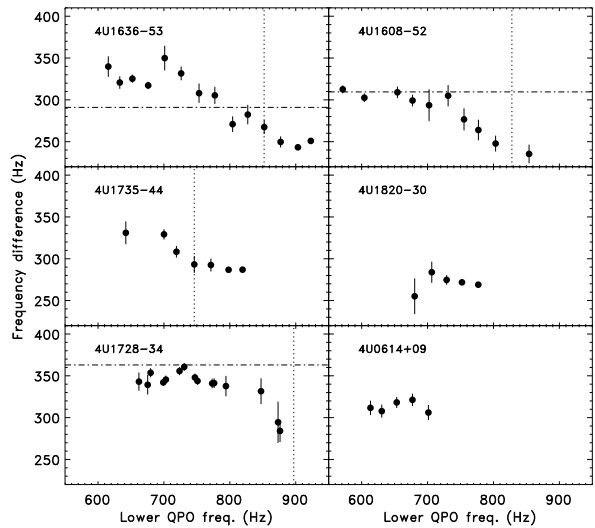}
\caption[]{Frequency difference versus lower QPO frequency. When
available the inferred spin frequency or half the spin frequency
is shown with a dashed line.
The frequency at which the quality factor of the lower QPO
frequency reaches its maximum value is also shown as a vertical
line.}
\label{barret_f3}
\end{figure}

\section{Discussion}
All current QPO models face challenges in explaining the high
quality factor observed, its dependence on frequency, and the
significantly different behaviours of the quality factors of the upper
and lower QPOs as reported here. It is therefore difficult to be
certain of the implications of our results.  However, as we discussed
in Barret et al. (2005a,b), the evidence in 4U~1636--536 for a
frequency ceiling independent of count rate and a sharp drop in the
quality factor at a fixed frequency is consistent with prior
expectations for effects due to the ISCO.  Such an inference would be
important enough that it is essential to determine the implications in
terms of neutron star masses, to examine alternate interpretations, and
also to determine if the current data are at least somewhat
quantitatively consistent with simple models involving the ISCO.

\subsection{Implication for neutron star masses}
Under the ISCO interpretation one would expect different sources to show different frequencies of peak $Q$ and different frequencies at which $Q$ extrapolates to zero ($\nu_{\rm lower,Q=0}$). Using $\nu_{\rm lower,Q=0}$ to infer the ISCO frequencies, on should then verify that the latter are compatible with our understanding of neutron star masses and evolution. 

Figures~2 shows that the first two criteria are indeed met.  Where it can be determined, the peak $Q$  frequency ranges from $\sim 750$~Hz in 4U~1735--44 to $\sim 900$~Hz in 4U~1728--34. 
Similarly, $\nu_{\rm lower,Q=0}$ ranges from $\sim 850$~Hz in
4U~1735--44 to $\sim 950$~Hz in 4U~1728--34.

Inferring the ISCO orbital frequency requires a somewhat more
specific model.  All models of neutron star kHz QPOs that can
accommodate the data suggest that the upper kHz QPO is close to an
orbital frequency (e.g., Miller et al. 1998; Lamb \& Miller 2001,
2003)  or a vertical epicyclic frequency (e.g., Stella \& Vietri
1998; Stella, Vietri, \& Morsink 1999; Psaltis \& Norman 2000;
Abramowicz et al. 2003;  Lee, Abramowicz, \& Klu\'zniak 2004; Bursa
et al. 2004; Klu\'zniak \& Abramowicz 2005), which deviates from an
orbital frequency by at most a few Hertz around a neutron star
(Markovi\'c 2000).  In such models, we can use the observation that
the separation between the upper and lower peaks is close to either
the spin frequency or half the spin frequency (e.g., see the panels
for 4U~1636--53, 4U~1608--52, and 4U~1728--34 in Figure~4) to
estimate the orbital frequency at the ISCO.  If we simply estimate
$\nu_{\rm ISCO}$ by adding half the spin frequency or the spin
frequency to $\nu_{\rm lower,Q=0}$, as appropriate, we find
$\nu_{\rm ISCO}\approx 1220$~Hz for 4U~1636--53, $\nu_{\rm
ISCO}\approx 1230$~Hz for 4U~1608--52, and $\nu_{\rm ISCO}\approx
1310$~Hz for 4U~1728--34, the three of our six sources with a known
spin frequency.  From these we can estimate the gravitational
masses  of the neutron stars (see Miller et al. 1998):
\begin{equation}
M\approx 2.2\,M_\odot(1000~{\rm Hz}/\nu_{\rm ISCO})(1+0.75j)
\end{equation}
where $j\equiv cJ/(GM^2)\sim 0.1-0.2$ is the dimensionless angular
momentum of the star.  The inferred masses of these sources are
therefore in the range of $\sim 1.8-2.1\,M_\odot$, which is higher
than inferred from radio timing of double neutron star binaries
(see, e.g., Cordes et al. 2004 for a recent review) but is
consistent with the higher masses derived for Vela X-1 (Quaintrell
et al. 2003) and some radio pulsars in binaries with detached
low-mass stars (Nice et al. 2005).  These masses are also compatible
with realistic modern equations of state, which predict maximum
masses for slowly rotating ($j\ll 1$) stars of $\sim
1.8-2.3\,M_\odot$ (Akmal, Pandharipande, \& Ravenhall 1998; Lattimer
\& Prakash 2001; Kl\"ahn et al. 2006).  To this level, then,
identification of the rapid drop in quality factor as being caused
by an approach to the ISCO is consistent with observations and
theoretical knowledge about neutron star structure.

\subsection{Role of the magnetic field}

There do not currently exist any specific suggestions about mechanisms
not involving the ISCO that would produce the observed drop in $Q$.
Nonetheless, we can follow Barret et al. (2005a,b) in speculating
that interaction with the stellar magnetic field may have an effect on the QPO coherence. This speculation may be supported by the fact that the magnetic field is likely to be involved in the QPO generation, through the role played by the spin in the setting of the QPO frequencies (e.g in the forced resonance model of Lee et al. (2004), the magnetic field may provide the necessary periodic perturbation of the disk at the neutron star spin frequency). The potentially complex magnetic field geometry,
especially near the star, means that it is difficult to establish
firm predictions about the behaviour of the quality factor.  However,
one would expect that, because quantities such as the Alfv\'en radius
and thickness of the disk depend on mass accretion rate, the frequency
at which the quality factor is maximal and the frequency at which it
extrapolates to zero would also depend on the mass accretion rate
if the magnetosphere is involved.
Evaluation of this expectation against the data is difficult because
there is no direct measure of the mass accretion rate. 

However, it is already worth noting that the quality factor is not well
correlated with the X-ray count rate, which may be considered as a
proxy for the mass accretion rate, as shown in Figure 5 in the case
of 4U~1636--536 (the same behaviour is observed for other sources, Barret et al. 2005b). 
Although this does not absolutely rule out interpretations based on
plasma physics, the fundamental dependence on the frequency is more
naturally  understood if it is based on spacetime properties such as
the ISCO.
\begin{figure} 
\includegraphics[width=0.45\textwidth]{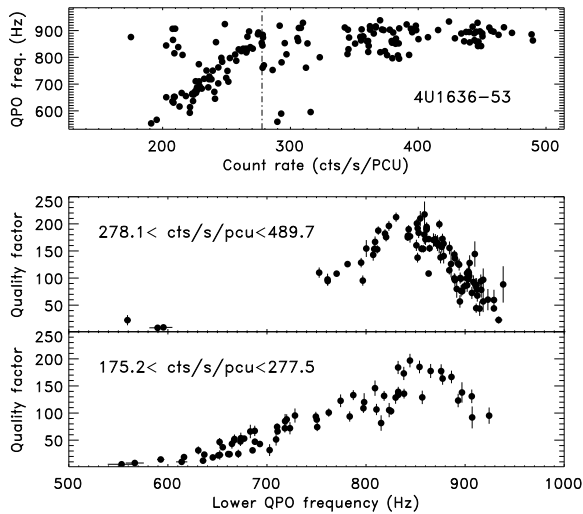}
\caption[]{(top panel) The frequency ceiling of the lower 
kHz QPO of 4U~1636--536.
(bottom panel) The behaviour of the quality factor in two count rate
regimes, delimited by the mean count rate over the whole observing set.
The count rates are normalized per RXTE/PCA/PCU units and are estimated
over the full energy range covered by the PCA.}
\label{barret_f4}
\end{figure}

\subsection{Advection based toy model}

In this section we describe a toy model that suggests how
the quality factor changes as the region of the QPO generation approaches
the ISCO.  Generically, we suppose that the frequency of the QPOs is determined by some mechanism in an active region (at $r_{\rm orb}$) of finite extent (width $\Delta r_{\rm orb}$) in the disk.  Note that on energetic grounds the emission itself is likely to be produced at the stellar surface, even if the frequency is determined in the disk. The oscillating phenomenon
has a characteristic lifetime (i.e. lasts a certain
number of cycles, $N_{\rm cycles}$). It is further assumed to be linked to
the accreting matter, hence it is advected through the active region
with a radial velocity $v_{r}$.  As we discuss in more
detail below, there are then three basic parts of the
observed width of the QPO: (1)~a contribution from the finite
extent $\Delta r_{\rm orb}$, (2)~a contribution from the radial
drift $\Delta r_{\rm drift}$ during the lifetime of the oscillation,
and (3)~a contribution from the finite lifetime itself.

We now present a simplified model for these contributions. It is worth stressing that the model discussed below does not provide a physical explanation for the quality factor of QPOs, it simply shows how, under some reasonable assumptions, the above contributions may combine to give its observed frequency dependence, as found in several sources. Our prime focus in this paper is the
sharp drop of $Q$ for the lower peak at its highest observed
frequencies.  We therefore concentrate on this first, showing
that the flattening of specific angular momentum curve close
to but outside the ISCO naturally leads to the observed 
high-frequency behaviour, even in our simplified mathematical 
formulation.  We then show that even if the contributions to the
QPO widths from drift and finite lifetime are kept fixed,
simple and physically reasonable modifications to just the 
contribution from $\Delta r_{\rm orb}$ are able to reproduce
both the quality factor behaviour of the upper peak frequency,
and the lower-frequency behaviour of the lower peak.  Therefore,
although our model is only illustrative, it demonstrates that
(1)~approach to the ISCO inevitably leads to a sharp drop in
$Q$ in a wide variety of models, and (2)~the general picture
of QPOs proposed by multiple authors is also consistent
with the full frequency dependence of the quality factor of
both peaks.

\subsubsection{Modeling the drop of the quality factor of the lower QPO}
\label{toy_lower}
Suppose that the lower peak is produced by the
interaction of either an orbital (Miller et al. 1998; Lamb \&
Miller 2001, 2003, at $\nu_{\rm orb}$) or a vertical epicyclic
(Stella \& Vietri 1998; Stella et al. 1999; Psaltis \& Norman
2000; Abramowicz et al. 2003; Lee et al. 2004; Bursa et al. 2004;
Klu\'zniak \& Abramowicz 2005, at $\nu_{\rm vert} \sim \nu_{\rm
orb}$) frequency at $r_{\rm orb}$ with the neutron star spin.  In this
picture, $\nu_{\rm lower}\approx
(\nu_{\rm orb}{\rm\ or\ }\nu_{\rm vert})-(\nu_{\rm spin}{\rm\ or\ }
\nu_{\rm spin}/2)\approx \nu_{\rm orb}-(\nu_{\rm spin}{\rm\ or\ }
\nu_{\rm spin}/2)$.   

If we assume that the interaction with the spin is resonant, the
resonance is not expected to add significantly to the breadth,
thus the frequency width of the lower peak is influenced by the
following factors:
\begin{enumerate}
\item $\Delta\nu_{\rm drift}$ accounting for the change of frequency
over the lifetime of the oscillating phenomenon as the gas advects in

\item $\Delta\nu_{\rm orb}$ corresponding to the range of available
orbital frequencies within the active region

\item  $\Delta\nu_{\rm life}$ produced by the finite lifetime of the
oscillating phenomenon
\end{enumerate}

As the true distribution of frequencies within $\Delta\nu_{\rm orb}$ and the true waveform of the oscillations are unknown (leading to uncertainties also on $\Delta\nu_{\rm drift}$), we approximate the total frequency width with the quadratic sum of the three contributions:
\begin{center}
\begin{equation}
\Delta\nu_{\rm total}=\sqrt{(\Delta\nu_{\rm drift})^2+
(\Delta\nu_{\rm orb})^2+(\Delta\nu_{\rm life})^2}\; ,
\end{equation}
\end{center}
and the quality factor is $Q_{\rm lower}=
\nu_{\rm lower}/\Delta\nu_{\rm total}$. In the following, we further assume that there are no systematics in our effort to track the QPO, so that the shift-and-add technique retrieves the intrinsic width $\Delta\nu_{\rm total}$ of the QPO. We now estimate 
each of these contributions in turn.

{\it Estimate of $\Delta\nu_{\rm drift}$}---In order
to produce a quantitative model without too many free
parameters, let us consider test particle orbits in a Schwarzschild
spacetime; we expect that spacetimes with moderate spin ($j<0.3$)
will have similar behaviour. In such a spacetime, the specific
angular momentum of a fluid element is $\ell=M^{1/2}r(r-3M)^{-1/2}$,
where we use geometrised units in which $G=c\equiv 1$.  Loss of
angular momentum will therefore lead to inward radial motion and
thus a change in the orbit.  In particular, if the orbital period
$P_{\rm orb}=2\pi(r^3/M)^{1/2}$ is roughly constant during the
lifetime of the oscillating phenomenon (as it must be for the
observed $Q>10$), then for an average specific angular  momentum
loss rate $\langle{\dot\ell}\rangle$ the total radial drift during
$N_{\rm cycle}$ cycles is
\begin{equation}
\begin{array}{rl}
\Delta r_{\rm drift}&=2\pi N_{\rm cycle}(r^3/M)^{1/2}{\dot r}\\
&=2\pi N_{\rm cycle}(r^3/M)^{1/2}{2(r-3M)^{3/2}\over{M^{1/2}
(r-6M)}}\langle{\dot\ell}\rangle\; .\\
\end{array}
\end{equation}
As expected, close to the ISCO ($r=6M$ for a Schwarzschild geometry) where
the specific angular momentum decreases slowly with decreasing
radius, loss of angular momentum produces rapid radial drift.
The contribution of this drift to the width of the QPO is then
\begin{equation}
\Delta\nu_{\rm drift}\approx {3\over 2}{\Delta r_{\rm drift}\over{
r_{\rm orb}}}\nu_{\rm orb}
\end{equation}

{\it Estimate of $\Delta\nu_{\rm orb}$}---The width
$\Delta\nu_{\rm orb}$ is set by the radial extent of the active region;
this is the most difficult parameter to estimate in this toy model. We
will first rely on a somewhat specific model to get a first estimate of
$\Delta\nu_{\rm orb}$, then we will discuss the implications of our
assumption.

For the model specific estimate we will assume that the active region
is located at the inner edge of the disk where the radial velocity of
the accreting gas increases dramatically, for example in response to
an efficient removal of angular momentum by the stellar radiation
field or stellar magnetic field (e.g. Miller et al. 1998). We will
assume that the extent of the active region corresponds to an
increase of the radial speed comparable to the initial radial speed.
With this assumption, we obtain:

\begin{equation}
\Delta r_{\rm orb}={v_r(r_{\rm orb})\over{d{\dot\ell}/dr}}
{(r-6M)M^{1/2}\over{2(r-3M)^{3/2}}}\; ,
\end{equation}
and the corresponding frequency width is
\begin{equation}
\Delta\nu_{\rm orb}\approx{3\over 2}{\Delta r_{\rm orb}\over{
r_{\rm orb}}}\nu_{\rm orb}
\end{equation}

{\it Estimate of $\Delta\nu_{\rm life}$}---The final 
contribution is simply $\Delta\nu_{\rm life} 
=\nu_{\rm lower}/(\pi N_{\rm cycle})$.  In the
following we will assume that the number of cycles is
constant over the frequency range considered.

{\it Comparison with data}---This model of quality 
factor versus frequency depends on
four parameters: $N_{\rm cycle}$, $v_r(r_{\rm
orb})/(d{\dot\ell}/dr)$, $\langle{\dot\ell}\rangle$, and $\nu_{\rm
ISCO}$.  Assuming for simplicity that all these parameters are
constant for a given source, we show in Figure 6 a comparison of a
by-eye fit to the data for 4U~1636--536 with parameter
values of 100, 0.8, $3\times 10^{-8}$, and 1220~Hz.  Note that
the value of $\langle{\dot\ell}\rangle$, although apparently
small, simply corresponds to the value that would produce a
$\sim$1\% radial drift during the 100 cycle lifetime of the QPO.  

\begin{figure} 
\includegraphics[width=0.45\textwidth]{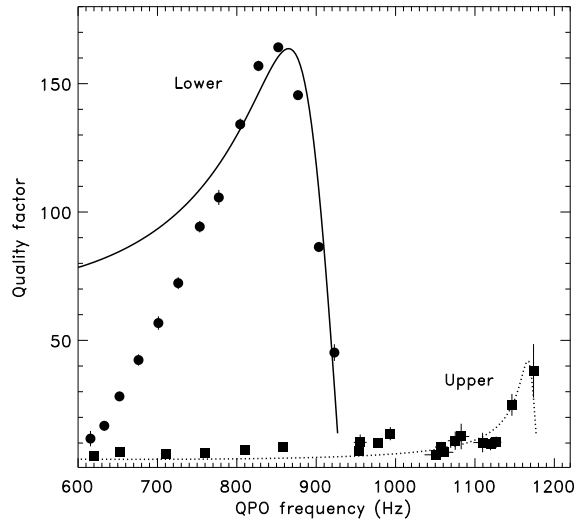}
\caption[]{Measured quality factor of the lower and upper
kHz QPOs of 4U~1636--536, compared to the quality
factor estimated with the toy model for the lower (solid
line, see section \ref{toy_lower}) and for the upper
(dashed line, see section \ref{toy_mod}) QPOs.}
\label{barret_f5}
\end{figure}

For these reasonable parameter values, the toy model reproduces
quantitatively the abrupt drop of the quality factor at high
frequencies. This is the region where the main contribution to the
QPO width is $\Delta\nu_{\rm drift}$. The simple toy model also
reproduces the peak in the quality factor.  These are generic
features of the flattening of specific angular momentum outside
the ISCO: the active region narrows, and the drift speed increases,
closer to the ISCO.  The observed behaviour at high frequencies
is therefore well-matched by these generic expectations.

At lower frequencies, however, $Q$ for the lower peak is much
lower than expected in this highly simplified model.  In addition,
the quality factor for the upper peak is typically much lower than for the
lower peak when both are seen together.  Although these are not
the main focus of our analysis, we now evaluate whether
plausible extensions to our picture can also match these
other behaviours quantitatively.

\subsubsection{Modification of the toy model}
\label{toy_mod}
As a guide to the necessary modifications, we show in Figure~7 the relative
contribution of the different factors to the total width of the
QPO. This figure shows that in our model, $\Delta\nu_{\rm orb}$ is
the dominant contributor
to the width for $\nu_{\rm lower}<800$ Hz. In order to
match the data, $\Delta\nu_{\rm orb}$ should keep increasing with
decreasing frequency (it should be $\sim 60$ at 600 Hz whereas our
model predicts $\sim 7$).  

\begin{figure} 
\includegraphics[width=0.45\textwidth]{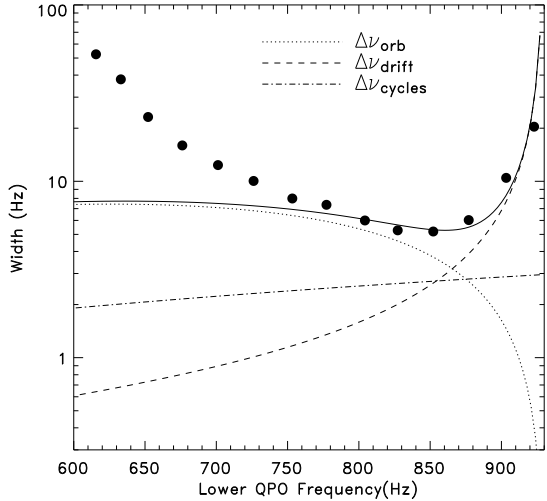}
\caption[]{Different contributions to the total QPO width (solid line)
estimated in the toy model, compared with the measured width for the lower
QPO of 4U~1636--536. The three contributions are shown:
$\Delta\nu_{\rm drift}$ (long dashed line), accounting for the
change of frequency over the lifetime of the oscillating phenomenon
as the gas advects inwards; $\Delta\nu_{\rm orb}$ (short dashed line),
corresponding to the range of available orbital frequencies within
the active region; and $\Delta\nu_{\rm life}$ (dotted line), produced by
the finite lifetime of the oscillating phenomenon.
}
\label{barret_f7}
\end{figure}
Therefore, in the framework of the toy model discussed above, we can keep
the same parameters $\Delta\nu_{\rm drift}$ and $\Delta\nu_{\rm life}$, but adjust 
$\Delta\nu_{\rm orb}$, so that a good match
to the data is obtained. This is shown in Figure 8 where the
difference between the observed data points and the contribution of
$\Delta\nu_{\rm life}$ and $\Delta\nu_{\rm drift}$ (i.e.
$\Delta\nu_{\rm orb}$) to the width has been approximated by a power
law. From $\Delta\nu_{\rm orb}$, one can derive an estimate from the
data of the extent of the active region (i.e. $\Delta r_{\rm
orb}/r_{\rm orb}$). The bottom of Figure 8 indicates that the size
of the active region, which determines the shape of the Q-frequency
curve at the low frequency end, is relatively broad at low frequencies
but decreases rapidly at higher frequencies.

With the model described above, it is tempting to investigate which modification would be needed to account for the very different behavior of the upper peak. For the upper peak we note that $Q$ tends to be much
lower than for the lower peak, that for 4U~1636--536 there is an
abrupt rise in $Q_{\rm upper}$ at the highest frequencies, and
that there is no visible drop.  Remarkably, all these features
can be reproduced just by modifying our simple model by
\begin{equation}
\Delta\nu_{\rm orb,upper}=x_{\rm upper}\Delta\nu_{\rm orb,lower}
\end{equation} 
where $x_{\rm upper}$ is a free parameter that is assumed independent
of radius for simplicity, and the other contributions are kept fixed.
As Figure~6 shows, $x_{\rm upper}=20$ matches the data for 
4U~1636--536 surprisingly well, where $\nu_{\rm upper}$ is computed
for a given $\nu_{\rm lower}$ using the observed frequency separation.  
Note that the increased breadth
means that $\Delta\nu_{\rm drift}$ and $\Delta\nu_{\rm life}$ only
become important closer to the ISCO, leading to a sharper drop in
$Q$ that would be difficult to detect due to the low amplitude of
the oscillation and narrow frequency range in which it is visible.

Is it plausible that $\Delta\nu_{\rm orb}$ would behave as needed
to produce the observed quality factors?  Although no theory currently
exists to provide quantitative predictions, we can argue that this
behaviour is reasonable.  For example, the upper peak is known to
have different properties than the lower (e.g., see Barret et al. 2005a),
so additional broadening mechanisms are to be expected.  One specific
proposal, made by Lamb \& Miller (2001), is that if the upper peak
is produced by the movement of the footprints of gas accretion onto
the neutron star, then a contribution to the breadth and centroid
frequency of the QPO is the total azimuthal angle (and its time
derivative) traveled by gas from the active region to the stellar
surface.  This could be a major source of broadening, which would
plausibly dominate the total width and would be proportional to
$\Delta r_{\rm orb}$.  Similarly, if the width of the active region
is determined by angular momentum removal by radiation or magnetic
fields, then regions more distant from the star could easily be
screened more effectively, leading to a broader region and thus
lower quality factor QPOs.  Therefore, although we must wait for
full numerical simulations to get first-principles information about
neutron star QPO quality factors, the qualitative behaviour of both
QPOs is understandable in the current pictures as long as the
high-frequency drop in $Q$ is related to the ISCO.

 \begin{figure} 
\includegraphics[width=0.45\textwidth]{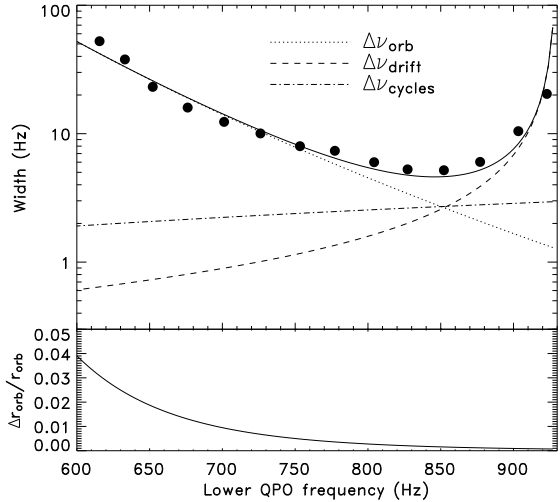}
\caption[]{Same as Figure 7, but with  $\Delta\nu_{\rm orb}$ fitted
by a power law to account for the rise of the QPO width towards
lower frequency (see section 3.3.2) (top).  $\Delta r_{\rm orb}/r_{\rm orb}$ as derived from
$\Delta\nu_{\rm orb}$ fitted above (bottom).}
\label{barret_f8}
\end{figure}

\section{Conclusions}

We conclude that current RXTE data are qualitatively and
quantitatively consistent with an ISCO-induced sharp drop in quality
factor in several low luminosity neutron star binaries.  If this is the correct
interpretation, this is a signature of an effect unique to strong
gravity, and it also provides strong constraints on models of matter
beyond nuclear density by requiring that neutron star gravitational
masses can exceed $2\,M_\odot$.  The most important observational
work now remaining is to determine whether models not involving the
ISCO are plausible.  The next step is thus to study the quality factor
against estimates of the mass rate such as count rate or hard or soft
color, and more generally spectral parameters. This will be the scope
of a forthcoming paper, using the unique properties of 4U~1636--536,
which stands currently as the best candidate for the ISCO signature.

\section{Acknowledgements}
We thank Mariano M\'endez and Michiel van der Klis for extensive
discussions, and Jean-Pierre Lasota for comments on the paper.  MCM was supported in part by a senior NRC
fellowship at Goddard Space Flight Center.  This research has
made use of data obtained from the High Energy Astrophysics
Science Archive Research Center (HEASARC), provided by NASA's
Goddard Space Flight Center. We are grateful to an anonymous referee for her/his comments that helped to strengthen some of the points presented in this paper.

\end{document}